\documentclass[aps,prb,reprint,superscriptaddress]{revtex4-1}

\usepackage{amsmath}    
\usepackage{gensymb}
\usepackage{graphicx}   
\usepackage[final]{changes}
\begin{document}

\title{
Vector magnetometry using silicon vacancies in 4H-SiC at ambient conditions}

\author{Matthias Niethammer}
\author{Matthias Widmann}
\author{Sang-Yun Lee}
\email{s.lee@physik.uni-stuttgart.de}

\affiliation{3rd Institute of Physics, University of Stuttgart, Pfaffenwaldring 57, 70569 Stuttgart, Germany}

\author{Pontus Stenberg}
\author{Olof Kordina}
\affiliation{Department of Physics, Chemistry and Biology, Link\"{o}ping University, SE-58183 Link\"{o}ping, Sweden}

\author{Takeshi Ohshima}
\affiliation{National Institutes for Quantum and Radiological Science and Technology, Takasaki, Gunma 370-1292, Japan}

\author{Nguyen Tien Son}
\affiliation{Department of Physics, Chemistry and Biology, Link\"{o}ping University, SE-58183 Link\"{o}ping, Sweden}

\author{Erik Janz\'{e}n}
\affiliation{Department of Physics, Chemistry and Biology, Link\"{o}ping University, SE-58183 Link\"{o}ping, Sweden}

\author{J\"org Wrachtrup}
\affiliation{3rd Institute of Physics, University of Stuttgart, Pfaffenwaldring 57, 70569 Stuttgart, Germany}
\affiliation{Max Planck Institute for Solid State Research, Heisenbergstra{\ss}e 1, 70569 Stuttgart, Germany}

\date{\today}

\begin{abstract}
Point defects in solids promise precise measurements of various quantities. Especially magnetic field sensing using the spin of point defects has been of great interest recently. When optical readout of spin states is used, point defects achieve optical magnetic imaging with high spatial resolution at ambient conditions. Here, we demonstrate that genuine optical vector magnetometry can be realized using the silicon vacancy in SiC, which has an uncommon S=3/2 spin. To this end, we develop and experimentally test sensing protocols based on a reference field approach combined with multi frequency spin excitation.
Our works suggest that the silicon vacancy in an industry-friendly platform, SiC, has potential for various magnetometry applications at ambient conditions.




\end{abstract}


\maketitle

\section{Introduction}
In the past decade, quantum magnetometry based on atomic scale defects such as the nitrogen-vacancy (NV) centers in diamond has attracted considerable interest since it can be utilized in various applications ranging from material to life sciences~\cite{Schirhagl2014,LeSage2013,McGuinness2011,Hall2013,Hall2012,Hall2010}.
The NV high spin system (S=1) and its C$_{\textnormal{3V}}$ symmetry allows determining not only the field strength but also the polar angle orientation of the external magnetic field~\cite{Balasubramanian2008,Steinert2010}.
The long-lived spin states and optically detected magnetic resonance (ODMR) have led to high sensitivity~\cite{Schirhagl2014}
and when combined with optical or scanning probe microscopy, optical magnetic imaging with nanometer scale spatial resolution has been demonstrated as well~\cite{Balasubramanian2008,Degen2008a,Maertz2010,Hall2013,Hong2013,Grinolds2014,2016arXiv160202948J}.

Recently, silicon carbide (SiC) has been recognized as an emerging quantum material potentially offering a platform for room temperature wafer scale quantum technologies~\cite{Weber2010,Koehl2011,Kraus2014,Kraus2014SR,Lohrmann2015,Widmann2015,Castelletto2014,Soltamov2015}, benefiting from advanced fabrication~\cite{Kimoto2014,Baliga2005,Sarro2000,Horsfall2007}. Many intrinsic defects\deleted{ exist similar to diamond}, and their optical and spin-related properties vary depending on the polytype~\cite{Falk2013}. Among them, the divacancy and silicon vacancy (V$_{\textnormal{Si}}$) in hexagonal and rhombic polytype SiC are known to have a spin angular momentum S$\mathrm{>}$1/2~\cite{Wimbauer1997,Mizuochi2002,Soltamov2015,Koehl2011,Falk2013}.
It has been recently shown that their spins are controllable and optically detectable on a single spin level at both room~\cite{Widmann2015} and cryogenic temperature~\cite{Christle2015} with a long spin coherence time~\cite{Yang2014,Widmann2015,Christle2015}.

High spin systems ($\mathrm{S>1/2}$) with a non-zero zero-field splitting (ZFS) in general allow for vector magnetometry because spin resonance transition frequencies depend on both strength and orientation of the applied magnetic field even when the Land$\mathrm{\acute{e}}$ g-factor is isotropic~\cite{Lee2015,Balasubramanian2008}. However, only partial orientation information can be extracted for spin systems with uniaxial symmetry as spin transition frequencies do not show azimuthal dependence~\cite{Lee2015,Balasubramanian2008}.
Therefore one is limited to sense only inclination or amplitude~\cite{Balasubramanian2008,Steinert2010,Lee2015,Simin2015}. Both the NV center in diamond and V$_{\textnormal{Si}}$ in hexagonal polytypes, e.g. 4H- and 6H-SiC, and a rhombic polytype, e.g. 15R-SiC, have the C$_{\textnormal{3V}}$ uniaxial symmetry, thus only allow to detect the polar angle of the applied field~\cite{Balasubramanian2008,Steinert2010,Lee2015,Simin2015}.
The four different NV orientations in diamond allow full reconstruction of field vectors, but it requires one to discriminate up to 24 possible orientations since one cannot find which transition belongs to which orientation~\cite{Steinert2010}. In order to circumvent this problem, one must apply reference fields~\cite{Steinert2010,Maertz2010}.
The C$_{\textnormal{3V}}$ symmetry and the single preferential spin orientation of the V$_{\textnormal{Si}}$ in SiC hinder genuine vector magnetometry since only the polar angle can be obtained~\cite{Lee2015,Simin2015}. However, the preferential alignment allows an unambiguous assignment of the observed resonance transitions
while overlap of several resonance transitions from different NV orientations~\cite{Lai2009,Alegre2007,Matsuzaki2016} adds complexity in experiments~\cite{Dmitriev2016} and limits precision of sensing.
This is a considerable advantage to cubic lattice systems such as diamond, where only complex growth can yield a similarly unique orientation~\cite{Michl2014,Lesik2014,Lesik2015,Pham2012a}.
Here we demonstrate that although the $\mathrm{V_{Si}}$ in 4H-SiC exhibits only a unique spin orientation with uniaxial symmetry, all vector components of a magnetic field can also be reconstructed by combining reference fields with multi frequency spin excitation.
Furthermore, the ZFS of the V$_{\textnormal{Si}}$ in hexagonal polytypes of SiC exhibits a very weak temperature dependence~\cite{Kraus2014SR}. These make the V$_{\textnormal{Si}}$ in SiC promising for magnetometry applications.


Below, we demonstrate how optical DC vector magnetometry can yield unambiguous measurement of the vector components of a magnetic field using the V$_{\textnormal{Si}}$ in one of the hexagonal polytypes, 4H-SiC.
We develop a simple model to explain transient spin excitation and the optical detection of spin signals. Their analysis provides a better understanding for the underlying optical cycle responsible for the ODMR of the V$_{\textnormal{Si}}$ in 4H-SiC.

\section{Electron spin resonance of silicon vacancy in silicon carbide}
\begin{figure}
\includegraphics[width=1\columnwidth]{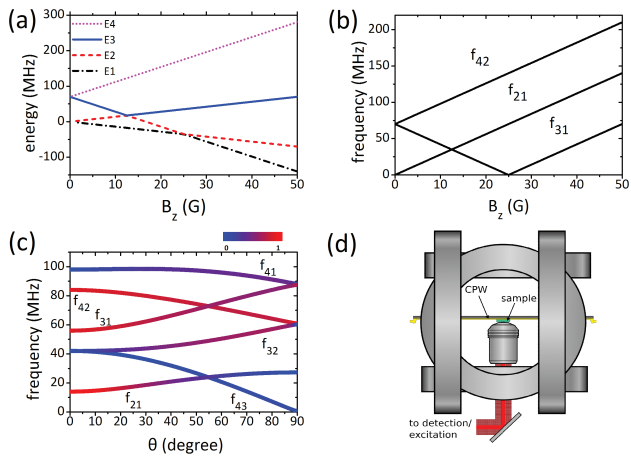}
\caption{\label{fig:B0dependence}(Color online) 
(a) Magnetic field strength dependence of the ground state spin quartet sublevels for V$_{\textnormal{Si}}$ in 4H-SiC  with S=3/2 when $B_{0}$ is aligned to the spin orientation. $2D$=70 MHz is assumed. E1, E2, E3, and E4 states are sorted by energy eigenvalues in ascending order and correspond to $M_{S}=-1/2,\, +1/2,\, -3/2,\, \mathrm{and}\, +3/2$ for $|B_{0}|<D$ and $M_{S}=-3/2,\, -1/2,\, +1/2,\, \mathrm{and}\, -3/2$ for $|B_{0}|>2D$. Energy is shown in frequency unit ($E=hf$).
Expected (b) magnetic field dependence and (c) polar angle dependence of the resonance transition frequencies of the ground quartet state for the V$_{\textnormal{Si}}$ in 4H-SiC when $B_0 || \mathrm{c-axis}$ and $B_0=0.5\,\mathrm{mT}$, respectively.
$f_{ij}$ is the resonance frequency between Ei and Ej states.
$f_{ij}=(Ei-Ej)/h$.
The color scale indicates the calculated transition probability with $B_{1}$ perpendicular to the c-axis.
(d) a part of experimental setup showing the SiC sample attached to a coplanar waveguide, which was used for RF irradiation, surrounded by three Helmholtz coil pairs.
See Ref.\cite{SM} for the details of experimental methods.
}
\end{figure}

The V$_{\textnormal{Si}}$ in 4H-SiC is a negatively charged spin 3/2 defect consisting of a vacancy on a silicon site which exhibits a C$_\textnormal{3v}$ symmetry~\cite{Janzen2009}, known as V2 or $\mathrm{T_{V2}}$ centers in literature. The relevant spin Hamiltonian of the system~\cite{Atherton1993,Janzen2009}, assuming uniaxial symmetry, is given as
\begin{equation}\label{eq:Hamiltonian}
H=hD[ S_{z}^{2} - S(S+1)/3]+g \mu_B  \vec{B_{0}} \cdot \vec{S}
\end{equation}
where $h$ is the Planck constant, $g$ is the electron Land$\mathrm{\acute{e}}$ g-factor, (2.004~\cite{Sorman2000}), $\mu_B$ is the Bohr magneton, and $\vec{B_{0}}$ describes the external magnetic field. Coupling to nuclear spins is ignored since $\mathrm{^{29}Si}$, the most abundant nuclear spin in SiC~\cite{Mizuochi2002,Mizuochi2003}, is diluted in our sample~\cite{SM}.
$D$ describes the axial component of spin dipole-dipole interaction.
This is responsible for a splitting of $\mathrm{ZFS}=2D$ between $|M_{S}|=3/2$ and $|M_{S}|=1/2$ states at a zero magnetic field~\cite{Lee2015} as shown in Fig.~\ref{fig:B0dependence}(a).
It has been suggested that
optical excitation leads to spin polarization into the $M_{S}=\pm 1/2$ spin sublevels of the ground state due to spin-dependent intersystem crossing (ISC)~\cite{Baranov2011,Soltamov2012,Simin2015,Fuchs2015,Widmann2015,Soykal2016}. The fluorescence emission is brighter when the system is in one of the $M_{S}=\pm 3/2$ states which is the basis for optical detection of electron spin resonance~\cite{Baranov2011,Widmann2015}.
Soykal \textit{et al.} recently claimed opposite: $M_{S}=\pm 3/2$ states are preferentially occupied and fluorescence emission is brighter when the $M_{S}=\pm 1/2$ and $D$ is negative~\cite{Soykal2016}. However, we will keep the former model for convenience as both two opposing models can explain the observed results.

The magnetic field dependence of the energy eigenvalues of each spin quartet sublevel in the ground state is shown in Fig.~\ref{fig:B0dependence}(a).
There is only a single transition at $f=2D$ when no magnetic field is applied, where $f$ is the resonance frequency. This degeneracy is lifted by an external magnetic field giving rise to multiple transitions. The number of observable transitions varies depending on the magnetic field orientation as shown in Fig.~\ref{fig:B0dependence}(b) and (c). $f_{42}$ and $f_{31}$, corresponding to $M_{S}=+3/2\leftrightarrow M_{S}=+1/2$ and $M_{S}=-1/2\leftrightarrow M_{S}=-3/2$, respectively, for $|B_{0}|<D$ and $\theta =0$, are most dominant and well observable in every orientation~\cite{Simin2015,Lee2015}. $f_{21}$ is also an allowed transition between $M_{S}=+1/2$ and $M_{S}=-1/2$ at $\theta =0$, and its strong transition probability is maintained for large $\theta $. However, the optically induced polarization into $M_{S}=\pm 1/2$ states does not induce a population difference between these two states, thus its ODMR signal is not observable~\cite{Sorman2000,Isoya2008,Kraus2014,Widmann2015}. $f_{41}$ and $f_{32}$ are forbidden for $\theta =0$ since they correspond to a $\Delta M_{S}=2$ transition, but are easily detectable when $\theta \neq0$ and $B_{0}< 1\, \mathrm{mT}$~\cite{Simin2015}. These multiple transitions will be used to realize vector magnetometry as follows.


\section{Principle of the vector magnetometry}
In general, spin system magnetometery exploits the magnetic field dependence of spin resonance transition frequencies to reconstruct the magnetic field vector components.
This is often difficult as an observed transition structure is not unique for an applied field~\cite{Steinert2010}. Thus, reference fields, whose amplitude and orientation are known, are used to extract additional information~\cite{Steinert2010,Maertz2010}.
Similar to the NV center in diamond~\cite{Balasubramanian2008}, one can extract the applied field strength using a formula for S=3/2 quartet system when an unknown magnetic field vector is applied~\cite{Lee2015}, for example,
\begin{widetext}
\begin{align}
\label{eq:B0}
B_0 = \left[ \frac{h}{5g \mu_{B}} \{  ( \sqrt{3} f_{\mathrm{avg}}+f_{32})^2-f_{42}f_{31}-2(\sqrt{3}+1)f_{32}f_{\mathrm{avg}}-2D^2   \} \right]^{\frac{1}{2}}
\end{align}
\end{widetext}
where $f_{\mathrm{avg}}\equiv(f_{31}+f_{42})/2$ (see Fig.~\ref{fig:B0dependence}(c)). Note that similar formulas utilizing other transitions, e.g. $f_{41}$ instead of $f_{32}$, and a formula for $\mathrm{cos^{2}\theta}$ can also be found~\cite{Lee2015}. The formulas show that as long as one can find three resonance transitions, the applied magnetic field strength can be explicitly determined if the ZFS is known. In order to precisely determine the vector components of the unknown stray magnetic field, whose amplitude is
\begin{equation}
\label{eq:stray}
 \left| \vec{B}_{\mathrm{s}} \right|=\sqrt{B_{\mathrm{s,x}}^2+B_{\mathrm{s,y}}^2+B_{\mathrm{s,z}}^2},
\end{equation}
three subsequent ODMR measurements with different reference fields should be performed. If the applied reference fields are perpendicular to each other, we obtain
\begin{equation}
\label{eq:ref_fields}
 \left|\vec{B}_{\mathrm{s}}+\vec{B}_{\mathrm{ref, i}} \right|^2= (B_{\mathrm{s,i}}+B_{\mathrm{ref,i}})^2 + B^2_{\mathrm{s,j}} + B^2_{\mathrm{s,k}}\\
\end{equation}
with $i,j,k \in  \left\{ \left( x,y,z \right) \right\}$. 
Using eq.(\ref{eq:stray}) and (\ref{eq:ref_fields}),
\begin{equation}
\label{eq:final}
B_{\mathrm{s,i}}=\frac{\left|\vec{B}_{\mathrm{s}}+\vec{B}_{\mathrm{ref, i}}\right|^2 -\left|\vec{B}_{\mathrm{s}}\right|^2-B^2_{\mathrm{ref,i}}}{2 B_{\mathrm{ref,i}}}
\end{equation}
Therefore, all the vector components of the unknown stray field $B_{\mathrm{s,i}}$ can be obtained explicitly.

\section{Methods and Materials}
To demonstrate proof-of-principle experiments, we performed ODMR experiments without and with three reference fields
(see Fig.~\ref{fig:DCOffsetODMR}).
The sample used in the experiments was a $\mathrm{350\,\mu m}$ thick $\mathrm{^{28}Si}$ enriched 4H-SiC layer grown on a natural 4H-SiC substrate in a horizontal hot-wall chemical vapor deposition system~\cite{SM}.
The sample was irradiated by 2 MeV electrons with a dose of $\mathrm{10^{16} cm^{-2}}$ to create $\mathrm{V_{Si}}$ ensembles ($\mathrm{[V_{Si}]\simeq 2\times 10^{14} cm^{-3}}$)~\cite{SM}. In the ODMR experiments, the sample was excited with a 785 nm laser focused by a lens.
The fluorescence light from the sample was detected by a femtowatt Si photodiode or APDs after a 835 nm longpass filter. ODMR measurements were performed using a virtual lock-in for both continuous-wave and pulsed ODMR~\cite{SM}.
Reference fields were applied by three coil pairs in Helmholtz configuration
(see Fig.~\ref{fig:B0dependence} (d)).
The ZFS of the $\mathrm{V_{Si}}$ in this sample was calibrated by measuring the maximum splitting between two allowed transitions, $f_{42}$ and $f_{31}$ while applying $|B_{0}|\gg \mathrm{ZFS}$ around the c-axis (See Fig.~\ref{fig:B0dependence}(b) and (c)). The obtained ZFS ($2D$) was $69.99\pm 0.03\, \mathrm{MHz}$ (data not shown). All measurements were performed at room temperature.





\section{Experimental results}
\begin{figure}
\includegraphics[width=\columnwidth]{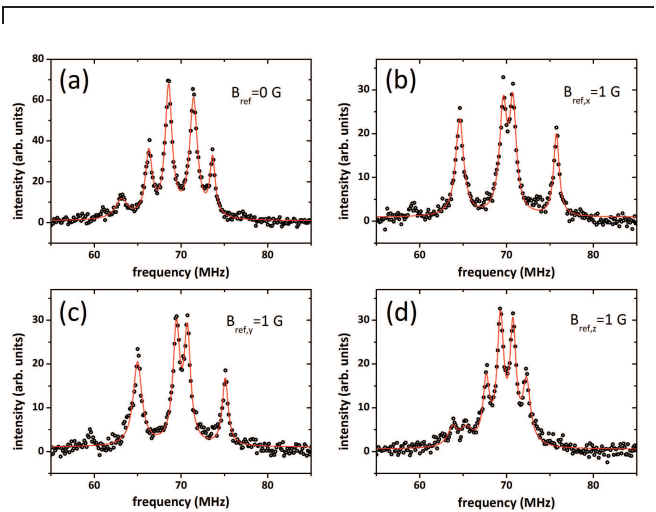}
\caption{\label{fig:DCOffsetODMR}(Color online) ODMR spectra without and with reference fields.
(a) ODMR spectrum without reference fields for which the stray magnetic field is to be determined. (b),(c), and (d) ODMR spectra with reference fields of $\left|B_0\right|=0.1\,\mathrm{mT}$ applied in x, y, and z directions, respectively. The red solid lines are the Lorentzian fit functions.
}
\end{figure}

The measured continuous-wave ODMR spectra for a zero applied field and three reference fields of $0.1\,\mathrm{mT}$ are depicted in Fig.~\ref{fig:DCOffsetODMR}. One can identify four transitions corresponding to $f_{41}$, $f_{42}$, $f_{31}$, and $f_{32}$ in all the observed spectra. It is, however, not possible to distinguish $f_{42}$ and $f_{31}$ using a single spectrum since their positions are interchanged at around the magic angle (see Fig.~\ref{fig:B0dependence}(c))~\cite{Lee2015}. Accurate field measurements are, however, still possible because only $f_{\mathrm{avg}}$ and $f_{42}\cdot f_{31}$ are necessary to calculate the applied magnetic field strength as seen from eq.(\ref{eq:B0}). Note that additional peaks of unknown origin appear $\sim5\,\textnormal{MHz}$ below the $f_{32}$ transition, which are under investigation and beyond the scope of this report.
Since only a single transition should be observable in the absence of any stray magnetic field, the four transitions in Fig.~\ref{fig:DCOffsetODMR}(a) obtained without an applied magnetic field indicate a stray magnetic field in the experimental environment. Applying equations (\ref{eq:B0}) and (\ref{eq:final}) to these data, we obtain the stray magnetic field vector components $B_{s,x}=0\pm3\,\mathrm{\mu T}$, $B_{s,y}=-18\pm3\,\mathrm{\mu T}$ and $B_{s,x}=-60\pm2\,\mathrm{\mu T}$.
These results were confirmed using a fluxgate sensor~\cite{SM}.

\begin{figure}
 \includegraphics[width=\columnwidth]{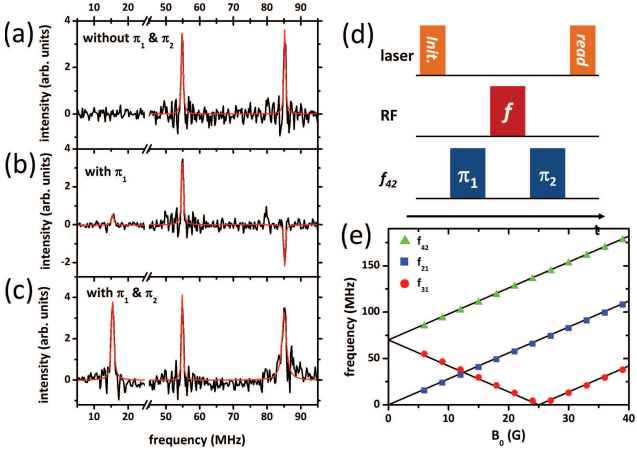}
 \caption{
 \label{fig:ELDOR}(Color online) Pulsed ODMR spectra at $B_{0}=0.543\pm0.003\,\mathrm{mT}$ and $\theta=3\pm1 \,\degree$.
 ODMR with (a) only a sweep pulse, (b) an additional $\pi$ pulse resonant to $f_{42}$ before the sweep pulse, (c) two additional $\pi$ pulses resonant to $f_{42}$ before and after the sweep pulse. The red solid lines are the Lorentzian fit functions. 
 (d) the used pulse sequences. A $600\,\mathrm{ns}$ long laser pulse is for optical spin polarization as an initialization pulse\deleted{, and readout too}.
 The same laser pulse is applied after the RF pulses for optical readout. A $\pi$ pulse whose frequency ($f$) is being swept from 5 MHz to 100 MHz is the sweep pulse. Two $\pi$ pulses resonant to $f_{42}$
 ($\pi_{1}$ and $\pi_{2}$)
 are used for swapping the populations.
 The overall length of each sequence is approximately $\mathrm{4\, \mu s}$.
 (e) frequencies of the three resonant transitions including $f_{21}$ obtained by the sequence used for (c) as a function of the applied magnetic field strength. The solid lines are the theoretical expectations. Error bars are smaller than the symbol size.}

 \end{figure}
The presented method based on ODMR with continuous-wave spin excitation is simple and allows an accurate field vector measurement.
However, since at least three transitions need to be visible, this method is not applicable under certain conditions. $f_{32}$ and $f_{41}$ become hardly detectable for a small polar angle~\cite{Lee2015,Simin2015}. Therefore, it is necessary to find a way to detect an additional allowed transition at $f_{21}$, which is usually not observable due to identical populations in $M_{s}=\pm 1/2$~\cite{Sorman2000,Isoya2008,Kraus2014,Widmann2015}.
One can create population difference between these two states by applying a $\pi$ pulse, swapping populations between $M_{s}= 3/2$ and $M_{s}= 1/2$ or $M_{s}= -1/2$ and $M_{s}= -3/2$~\cite{Isoya2008}.
As will be seen below, because a single population swapping between these two states does not allow to observe this hidden ODMR signal,
we investigated a few pulse sequences based on multi frequency spin excitation and established a \replaced{rate}{qualitative} model to explain how one can induce optical contrast of spin signals.


The pulse sequences and resulting ODMR spectra under $B_{0}=0.543\pm0.003\,\mathrm{mT}$ which was applied almost parallel to the spin sensor ($\theta = 3\pm1 \,\degree$) are compared in Fig.~\ref{fig:ELDOR}. Note that these values for the magnetic field strength and orientation were extracted from Fig.~\ref{fig:ELDOR}(c) using eq.(\ref{eq:B0}) and Ref.\cite{Lee2015}. These spectra exhibit additional side-peak structures because of excitation with a broad band rectangular RF pulse in contrast to the spectra in Fig.~\ref{fig:DCOffsetODMR} which were measured with continuous-wave spin and optical excitation. When a RF pulse, whose frequency was being swept from 10 to 100 MHz (sweep pulse), was used, only two allowed transitions, $f_{42}$ and $f_{31}$, were visible as shown in Fig.~\ref{fig:ELDOR}(a). Then, a $\pi$ pulse between $M_{S}=3/2$ and $M_{S}=1/2$ corresponding to $f_{42}$ was added before the sweep pulse in order to form a population difference between $M_{S}=\pm 1/2$ states. The missing transition $f_{21}$ was, however, very weak, and we detected negative signals at $f_{42}$ (Fig.~\ref{fig:ELDOR}(b)). When the same $\pi$ pulse was applied additionally after the sweep pulse, the $f_{21}$ transition was clearly visible with the other two transitions as well (Fig.~\ref{fig:ELDOR}(c)). In order to prove that this transition is from $f_{21}$, we monitored the magnetic field strength dependence of the three transition frequencies measured by the pulse sequence with two $\pi$ pulses at $f_{42}$ as shown in Fig.~\ref{fig:ELDOR}(e). The position of the $f_{21}$ transition is as expected from the spin Hamiltonian of eq.(\ref{eq:Hamiltonian})~\cite{Widmann2015}. Since the detected signal sign is ambiguous in the lock-in experiment~\cite{Lee2012}, we repeated these experiments without using lock-in methods and could confirm this result~\cite{SM}.



\begin{figure}
\includegraphics[width=1\columnwidth]{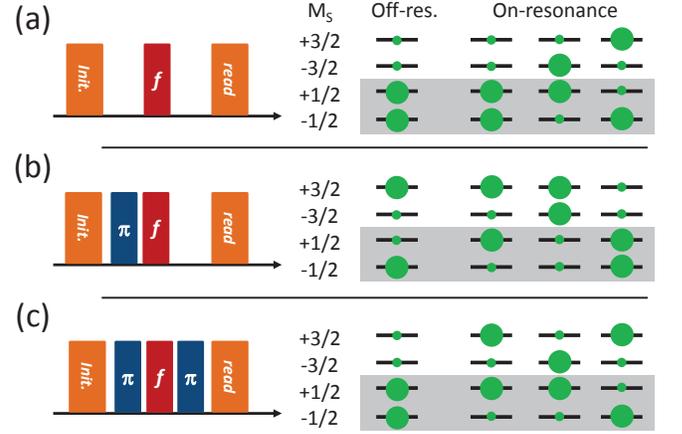}
\caption{\label{fig:rate}(Color online) Population redistribution by pulse sequences.
Left column: pulse sequences used for the spectra in Fig.\ref{fig:ELDOR} (a), (b), and (c), respectively. Right column: corresponding population distributions determined by each RF pulse sequence before the readout pulse. From left to right, when the sweep pulse is off-resonant and resonant at $f_{21}$, $f_{31}$, and $f_{42}$, respectively.}
\end{figure}
In order to explain the observed ODMR spectra in Fig.~\ref{fig:ELDOR}, we introduce a simplified model describing the ground state population redistributed by the used pulse sequences as depicted in Fig.~\ref{fig:rate}~\cite{SM}.
We find that the change in the fluorescence intensity by swapping populations between two states is either zero or $(b-d)\Delta n_{g,0}$. Here $d$ and $b$ are the rate related parameters of $|M_{S}|=1/2$ and $|M_{S}|=3/2$, respectively, whose difference is determined by only the difference in the ISC rates, and $\Delta n_{g,0}\equiv n_{d,g,0}-n_{b,g,0}$, where $n_{d,g,0}$ and $n_{b,g,0}$ are the initial population of a $|M_{S}|=1/2$ and $|M_{S}|=3/2$ state, respectively.
Since we assume that the $|M_{S}|=1/2$ states are highly populated by optical polarization and fluorescence emission is brighter when $|M_{S}|=3/2$ are highly occupied, $b>d$ and $\Delta n_{g,0}>0$. See Ref.~\cite{SM} for details. When a larger population is transferred to one of the $M_{S}=\pm 3/2$ states by the sweep pulse, one can see a fluorescence increase with respect to the off-resonance fluorescence intensity by $(b-d)\Delta n_{g,0}$, which is positive (see Fig.~\ref{fig:rate}(a)). This is consistent with the ODMR spectrum in Fig.~\ref{fig:ELDOR}(a).
When the sweep pulse follows a $\pi$ pulse at $f_{42}$,
one, two, and none of the $M_{S}=\pm 3/2$
states are highly populated at resonances by the sweep pulse at $f_{21}$, $f_{31}$, and $f_{42}$, respectively (see Fig.~\ref{fig:rate}(b)). Since $M_{S}=3/2$ is also highly populated when the sweep pulse is not resonant,
zero, $(b-d)\Delta n_{g,0}$, and $-(b-d)\Delta n_{g,0}$
at these frequencies will be observed.
This expectation is in agreement with what we experimentally observed as in Fig.~\ref{fig:ELDOR}(b).
Therefore, additional population swapping by a $\pi$ pulse at $f_{42}$ following the resonant sweep pulses will allow to have one of the $M_{S}=\pm 3/2$ states to be highly populated. In contrast, only the $M_{S}=\pm 1/2$ states will be highly populated at off-resonance as depicted in Fig.~\ref{fig:rate}(c), thus the same positive signals,
$(b-d)\Delta n_{g,0}$,
at the three resonances will appear. This is exactly equivalent to our experimental observations in Fig.~\ref{fig:ELDOR}(c).
This model can explain the signs and
relative
intensities of the observed ODMR signals well and detailed explanations
can be found in Ref.\cite{SM}.
We conclude that the presented sequence
as in Fig.~\ref{fig:rate}(c)
allows to observe the missing ODMR transition, and thus DC magnetometry becomes applicable for every orientation at the tested magnetic field strengths.

Though we aim to present proof-of-principle experiments for resolving an arbitrary magnetic field orientation, we provide discussions about the obtained sensitivity and its projection when the sample and detection methods are optimized. Note that if sensing the magnetic field strength is of only interest, phase detection methods, e.g. Ramsey interferometer, can be used instead which can enhance the sensitivity by many orders of magnitude~\cite{Degen2008}.
The sensitivity extracted from the ODMR spectrum in Fig.~\ref{fig:ELDOR}(c) using eq.(\ref{eq:B0}) and the formula for $\mathrm{cos^{2}\theta}$ in ref.\cite{Lee2015} is $0.2\, \mathrm{mT/\sqrt{Hz}}$ for the DC magnetic field strength and $30\, \mathrm{degree/\sqrt{Hz}}$ for the orientation. The number of $\mathrm{V_{Si}}$ of the used sample within the focal volume was quite small ($\sim2000$) since the confocal microscope with a high NA objective was used~\cite{SM}. If a larger $\mathrm{V_{Si}}$ concentration, e.g. $\mathrm{\sim10^{16}\, cm^{-3}}$~\cite{Fuchs2015}, is used, $30\, \mathrm{\mu T/\sqrt{Hz}}$ and $7\, \mathrm{degree/\sqrt{Hz}}$ can be expected with sub-wavelength spatial resolution. Substantial enhancement can be expected when high spatial resolution is not of interest; for example, up to $3\, \mathrm{nT/\sqrt{Hz}}$ and $0.002\, \mathrm{degree/\sqrt{Hz}}$ if $\mathrm{[\mathrm{V_{Si}}]\sim 10^{16}\, cm^{-3}}$ in a $\mathrm{1\,mm^{3}}$ volume device is used.
These sensitivities can be even further enhanced if optimum detection methods are used. For example, a light trapping waveguide and an optical cavity can improve the detection efficiency by many orders of magnitude~\cite{Clevenson2015,Dumeige2013}. A Hahn-echo sequence can be combined to the used sequence to improve the linewidth of the ODMR spectral lines. If a free precession time of $\mathrm{30\,\mu s}$ is used~\cite{Widmann2015,2016arXiv160205775S,Carter2015}, since linewidth of $\mathrm{\sim 10\, kHz}$ is expected, and the linewidth in Fig.3 is $\mathrm{\sim 500\, kHz}$, an order of magnitudes higher sensitivity is expected considering reduced duty cylce as well.

Now we discuss the dynamic range of the presented sensing methods.
For small magnetic fields, e.g. $B_{0} < {h D}/{g \mu_{B}}$, three transitions
are necessary.
Four transitions in Fig.~\ref{fig:DCOffsetODMR} have been successfully observed up to $0.8\, \mathrm{mT}$~\cite{Simin2015}. Thus, our methods are suitable for sub-mT DC vector magnetometry. When $B_{0} \sim {h 2D}/{g \mu_{B}}$, the suggested methods may not be useful because of complex spectra arising due to interactions among spin sublevels~\cite{Carter2015,He1993,2015arXiv151104663S}. At high magnetic fields, e.g. $B_{0} \sim$ 300 mT, two transitions $f_{21}$ and $f_{43}$ are well observable at every orientation as experimentally reported~\cite{Sorman2000,Kraus2014SR}. The forbidden transitions are hardly visible in high magnetic field ranges. Therefore, it should be further investigated whether the missing transition between $M_{S}=\pm 1/2$, which was successfully observed with multi frequency excitation for $B_{0}\parallel$ c-axis~\cite{Isoya2008} can be well observed independent on the field orientation in this field range.





\section{Summary}
We demonstrated DC vector magnetometry based on ODMR of S=3/2 quartet spins of the $\mathrm{V_{Si}}$ in 4H-SiC at room temperature. 
ODMR scans with reference fields realize reconstruction of all vector components of the unknown magnetic field.
We also demonstrated a pulse sequence based on multi frequency spin excitation as a complementary protocol
  to make this magnetometer practical. The suggested simple rate model also provides a better understanding for the optical cycle allowing ODMR. With this sensing protocol, very weak temperature dependence of the ZFS~\cite{Kraus2014} makes $\mathrm{V_{Si}}$ in SiC promising for robust magnetometer, and useful for optical magnetic imaging in nanoscale at ambient conditions. The possibility of electrically detected magnetic resonance~\cite{Umeda2012,Cochrane2012a,Bourgeois2015} in the wafer scale SiC may also allow for the construction of an integrated quantum device for vector magnetometry.


\begin{acknowledgments}
We acknowledge funding by the ERA.Net RUS Plus program (DIABASE), the DFG via priority programme 1601, the EU via ERC grant SQUTEC and Diadems, the Max Planck Society, the Knut and Alice Wallenberg Foundation, and KAKENHI (B) 26286047. We especially thank Corey Cochrane, Philipp Neumann, and Durga Dadari for inspiring discussions. We also thank Seoyoung Paik, Ilja Gerhardt, Florestan Ziem, Thomas Wolf, Amit Finkler, Roland Nagy, and Torsten Rendler for fruitful discussions.
\end{acknowledgments}


%

\end{document}